\documentclass[a4paper,usenatbib]{jpconf}
\usepackage{graphicx,amsmath,iopams}

\usepackage{color}

\newcommand{\xx}[1]{\!\times\!10^{#1}}
\newcommand{\PIC}{particle-in-cell}
\newcommand{\Eest}{E_\mathrm{est}}
\newcommand{\usc}{u_\mathrm{sc}}
\newcommand{\ush}{u_\mathrm{sh}}
\newcommand{\jcr}{j_\mathrm{cr}}
\newcommand{\Pcr}{P_\mathrm{cr}}
\newcommand{\Fcr}{F_\mathrm{cr}}
\newcommand{\Bls}{B_\mathrm{ls}}
\newcommand{\kres}{k_\mathrm{res}}
\newcommand{\Blsst}{B_\mathrm{ls,st}}
\newcommand{\Lss}{\lambda_\mathrm{ss}}
\newcommand{\LBst}{\lambda_{B,\mathrm{st}}}

\newcommand{\Beff}{B_\mathrm{eff}}
\newcommand{\Beffd}{B_\mathrm{eff,d}}
\newcommand{\pcc}{cm$^{-3}$}
\newcommand{\kmps}{km\,s$^{-1}$}
\newcommand{\rgz}{r_{g0}}
\newcommand{\xFEB}{x_\mathrm{FEB}}
\newcommand{\Reff}{R_\mathrm{eff}}
\newcommand{\Rtot}{R_\mathrm{tot}}
\newcommand{\Ez}{\mathbf{E}_0}

\newcommand{\Alf}{Alfv\'en}
\newcommand{\vscat}{v_\mathrm{scat}}

\begin{document}

\title{Monte Carlo modelling of  particle acceleration in collisionless shocks with an effective mean electric field}%

\author{S M Osipov$^{1}$, A M Bykov$^{1,2}$, D C Ellison$^{3}$}
\address{$^1$ Ioffe Institute, 26 Politekhnicheskaya st., St. Petersburg 194021, Russia}
\address {$^2$ Peter The Great St. Petersburg Polytechnic University, 29 Politekhnicheskaya st., St. Petersburg 195251, Russia}
\address {$^3$ North Carolina State University, Department of Physics, Raleigh, NC 27695-8202, USA}

\ead{osm2004@mail.ru}

\begin{abstract}
Relativistic particle acceleration in collisionless shocks of supernova remnants is accompanied by magnetic field amplification from cosmic ray (CR) driven plasma instabilities. Bell's fast CR-current instability is predicted to produce turbulence with a non-zero mean electric field in the shock precursor.
We present a Monte Carlo model of Fermi shock acceleration explicitly taking into account  an effective mean upstream electric field. Our model is nonlinear and includes the backreaction effects of efficient Fermi acceleration on the shock structure.\\
To be published in Journal of Physics: Conference Series (in print, 2019).
\end{abstract}

\section{Introduction}
The converging plasmas on  either side of a collisionless shock allow particle acceleration as particles scatter nearly elastically between the upstream and downstream regions.
The scattering centers are  magnetic field fluctuations moving  with the background plasma and multiple crossing of the collisionless shock front can result in significant energy  gains for individual particles.
%
%

This first-order Fermi mechanism [also called diffusive shock acceleration (DSA)] is expected to be efficient at the strong shocks in young supernova remnants (SNRs) (e.g. \cite{Ellison12}).
%
%
There is strong evidence to show that anisotropic CR distributions in the shock precursor leads to the development of plasma instabilities and magnetic field amplification (MFA) (e.g. \cite{VBK2005a}).
Furthermore, if the acceleration is efficient, the gradient of the CR pressure will cause a modification of the bulk flow of the background plasma in the upstream region.
The anisotropy of the CR distribution determines the growth of magnetic fluctuations in the precursor and the energy dependent scattering rate of fast particles. Since both of these depend critically on the bulk flow, DSA, with self-consistent MFA, is an essentially nonlinear problem which can only be fully addressed with computer simulations
(see e.g.  \cite{Jones91,Bell15}).
%

Several nonlinear models of efficient DSA in SNRs
were constructed to account for the backreaction of accelerated particles on the dynamics of the background plasma which carries the scattering centers   (see, for example, \cite{Caprioli2010,SchureEtal2012,BEOV14}).
Generally, the
kinetic description of fast particle transport and acceleration is
based on macroscopic recipes of particle scattering by magnetic turbulence.
Microscopic plasma instabilities  amplify magnetic fluctuations which are described after an appropriate averaging procedure on the ensemble of scattering centers. In the case of the resonant CR streaming instability \cite{Skilling1975,McKVlk82}, the
scattering centers  can move relative to the background plasma with some speed.

On the other hand, the short-scale modes produced
by the non-resonant Bell instability, which grow very fast in the shock precursor, have zero or small phase speed in comparison with the \Alf\ velocity \cite{Bell04}.
However,
 it was shown (see e.g. \cite{Bell04,Zirakashvili2008}) that  there is a mean, large-scale electric field directed opposite to the CR current producing Bell's instability.
This electric field is directed opposite to the CR current and modifies the  energy flux of Fermi accelerated particles in the shock precursor.
The effect of the mean electric field on the spectrum of shock accelerated particles was discussed in \cite{Zirakashvili2015} with a
test-particle approach using the  advection-diffusive equation with a mean electric field \cite{Fedorov1992}. Within this test-particle model, it was shown  that including the mean electric field resulted in a  decrease in the CR  proton acceleration efficiency.
Here,  we study the effect of the mean electric field on DSA with a nonlinear  Monte Carlo model which was developed in \cite{BEOV14}.
We show below  that nonlinear effects are essential for
modeling CR ions  with a
self-consistent mean electric field.  We don't consider CR electrons here since their precursor pressure is normally much less than that of ions and doesn't affect the flow dynamics.

\section{Monte Carlo model of DSA with electric fields}
We construct a steady-state model of a plane-parallel,
nonrelativistic collisionless shock where the nonlinear shock structure is determined iteratively. The model convergence is determined by satisfying the conservation of mass, momentum and  energy fluxes to some limit of accuracy.
%
Particles diffuse by pitch-angle scattering off magnetic field fluctuations (see \cite{BEOV14,Vladimirov09dis}).
The scattering is assumed to be isotropic and elastic in the local scattering center frame.

For CRs, the speed of the scattering centers, in the shock rest frame, is equal to the sum of the speed of the background plasma, $u(x)$, and the speed of the scattering centers relative to the background plasma $\vscat(x)$.
%
%
For the thermal background plasma, the scattering center speed is equal to the bulk plasma speed,  $u(x)$ (see \cite{EBJ96} for a full description of the pitch-angle diffusion process).
%
An important property of the Monte Carlo technique is that particle injection is included self-consistently once the scattering assumptions are made. A particle is ``injected" into the Fermi process if it is able to scatter back upstream having at least once crossed into the downstream region. Superthermal particles emerge smoothly from the background population and no artificial distinction between thermal particles and CRs is made.
%
There is no fluid approximation in the Monte Carlo method; the entire plasma, thermal and CR, is described as particles moving with a single set of scattering assumptions.

We begin our simulation with a mean electric field, $\Ez(x)$, at each spatial grid point. As particles scatter during time $dt$ (short compared to a collision time), they experience a momentum change from this field
$e \Ez dt$, where $e$ is the elementary charge (we consider the plasma consisting of protons only). This additional momentum change is included in our iterative scheme where we adjust the bulk flow speed $u(x)$, the background magnetic field turbulence, the overall compression ratio $\Rtot$, and the spatial dependence of the mean field $E_0(x)$, until mass, momentum, and energy fluxes are conserved throughout the shock.
%
The momentum and energy fluxes are calculated from the particle distribution functions and the magnetic field amplification is calculated self-consistently from the resonant and Bell's instabilities
as in \cite{BEOV14}. For simplicity, we do not include turbulent cascade here.

Ours is a plane-parallel model where all quantities depend only on the $x$-coordinate. The viscous shock is at $x=0$ and negative values of $x$ correspond to upstream values, i.e., the shock is directed along the $-x$-axis.
%
The nonperturbed far upstream magnetic field, $B_0$, and the average electric field are both directed parallel to the shock normal.
Particles leave the system by either convecting far downstream or by
escaping at an upstream free escape boundary (FEB), $\xFEB$.
For our results here, $B_0=3\,\mu$G, $\xFEB= -10^8\,\rgz$,
$\rgz \equiv m_p u_0 c/(eB_0)$, $u_0=5000$\,\kmps, $m_p$ is the proton mass, and $c$ is the speed of light.
%
%

The turbulent magnetic field is
$\displaystyle B_{w}\left(x\right)=\sqrt{4\pi\int_{0}^{\infty}W\left(x,k\right)dk}$,
where $W\left(x,k\right)$ is
the spectral energy density of the turbulence for wavenumber $k$,
$B_w(\xFEB)=B_0$,
and $\displaystyle W(\xFEB,k) \sim k^{-5/3}$.
The energy-containing scale is equal to 10 pc. The effective magnetic field is given by
$\displaystyle \Beff(x) = \sqrt{B_{w}^{2}\left(x\right)+B_{0}^{2}}$
(see \cite{Bykov11} for a fuller discussion of the turbulence calculation).

The position- and momentum-dependent mean free path of particles in the local scattering center frame is
\begin{equation}\label{lambda}
\lambda (x,p) =\frac{1}{\LBst(x,p)^{-1} + \Lss(x,p)^{-1}}
\ ,
\end{equation}
where
$\displaystyle\LBst(x,p) =\frac{pc}{e\Blsst(x,\kres)}$,
$\displaystyle\Lss(x,p) =\left(\frac{pc}{\pi e}\right)^{2}
\left [ {\int_{\kres}^{\infty}\frac{W\left(x,k\right)}{k}dk} \right]^{-1}$,
$\displaystyle \Bls(x,k)=
\sqrt{4\pi\int_{0}^{k}W(x,k) dk +B_{0}^{2}}$,
$\displaystyle \Blsst(x,k) =
\sqrt{4\pi\int_{0}^{k}W(x,k) dk}$,
%
$p$ is the particle momentum in the scattering center frame,
and
$\displaystyle \frac{\kres pc}{e \Bls(x,\kres)}=1$.

The rate of increase of the turbulent energy flux due to the CR driven plasma instabilities is equal to $\displaystyle\int_{0}^{\infty}\Gamma\left(x,k\right)W\left(x,k\right)dk$ , where $\displaystyle\Gamma\left(x,k\right)$ is the growth-rate of the plasma instability (see \cite{BEOV14}).
The change of the CR energy flux $\Fcr(x)$  due to CR interactions with the microscopic background plasma fields can be expressed as:
$\displaystyle \frac{d\Fcr(x)}{dx}= <\mathbf{J_{cr}}\mathbf{E}>$,
where the averaging of the product of microscopic CR current $\mathbf{J_{cr}}$, and the electric field  $\mathbf{E}$, is over scales larger than the CR scattering length.
In the macroscopic approach used in \cite{BEOV14,Zirakashvili2015}, as well as here,
$\displaystyle <\mathbf{J_{cr}}\mathbf{E}>$  can be expressed in two alternative ways.

First, it can be presented as
$\displaystyle <\mathbf{J_{cr}}\mathbf{E} >  =
\left[ u\left(x\right)+\vscat(x) \right ]
\frac{d \Pcr(x)}{dx}$,
where $\Pcr(x)$ is the CR pressure and the scattering center speed $\vscat$  can be derived self-consistently as described in \cite{BEOV14}.
Alternatively, the average
can be represented as in \cite{Zirakashvili2015}:
$\displaystyle <\mathbf{J_{cr}}\mathbf{E} >  =
u\left(x\right)\frac{d \Pcr(x)}{dx} +
\jcr(x) E_{0}\left(x\right)$,
where $\jcr(x)$ is the mean CR electric current in the local scattering center frame.
The first term on the right hand sides of
these equations
describes the adiabatic change of the CR energy flux.

The work done by CRs to  increase the turbulent energy flux due to
the CR-driven plasma instabilities is  represented in the macroscopic iterative model in two alternative ways.
One uses   $\vscat$ as a macroscopic parameter for the closure of the energy conservation equation, i.e.,
\begin{equation}\label{eq_v_scat}
\vscat(x) =
\frac{-\int_{0}^{\infty}\Gamma(x,k) W(x,k) dk}{d\Pcr(x)/dx}
\ .
\end{equation}
The other method uses the mean effective electric field,  i.e.,
\begin{equation}\label{eq_E0}
E_0(x) = \frac{-\int_{0}^{\infty}\Gamma(x,k)W(x,k)dk}{\jcr(x)}.
\end{equation}
Next we compare these two approaches  in the nonlinear Monte Carlo DSA model.

\section{Results}
We first perform test-particle calculations to  compare the effects of an average electric field (Eq.~\ref{eq_E0}) against one where the scattering center speed is determined with Eq.~(\ref{eq_v_scat}).
In test-particle models, in general, there is no modification of the upstream flow, i.e., $u(x)=\ush$, where $\ush$ is the shock speed measured in the frame of the far upstream  plasma.
We further assume constant values of the average electric field, $E_0$, the upstream scattering center speed, $\vscat$, and we ignore MFA.
We estimate values for $E_0$ and $\vscat$ that influence the CR spectrum in a similar fashion.
For this purpose, we equate changes of CR energy fluxes, i.e.,
$\jcr E_{0}= \vscat d \Pcr(x)/dx,$ for models with $E_0 \neq 0$ or $\vscat \neq 0$.
As definition of the CR current and pressure gradient, we use values 
from the model with
$\vscat \neq 0$  and $E_{0}=0$.

This model is equivalent to standard unmodified DSA \cite{bk88} with the scattering center speed in the upstream $u^{u}_{sc}=\ush + \vscat$ and in the downstream $\usc^d = \ush/\Rtot$.
In this case, the  upstream CR distribution function is
$\displaystyle f(x,p) = f_0(p) \exp{\left [\usc^u x/D(p) \right ]}$,
where $D(p) = \lambda(p) v/3$, $v$ is the particle speed,
and the upstream CR current is given by
$\displaystyle \jcr(x) =
-e \usc^u \int{f(x,p) d^3p}$.
By definition, $\displaystyle\Pcr(x) = \frac{1}{3} \int{v p f(x,p) d^3p}$,
and using the expression for $f(x,p)$, we obtain
$\displaystyle\frac{d\Pcr(x)}{dx} =
\usc^u \int{\frac{p}{\lambda(p)} f(x,p) d^3p}$.

In our test-particle calculations we assume $\lambda(p) \approx r_g$,
where $r_g = pc/(eB_0)$ is the gyro-radius of a particle with momentum $p$ in a magnetic field $B_{0}$.
We also  assume that the CR current doesn't change due to the action of the electric field.
With these approximations, the  estimated average electric field is
\begin{equation} \label{eq:Eest}
\Eest = - \vscat B_0/c
\ .
\end{equation}

\begin{figure}[h]
\begin{center}
\includegraphics[scale=0.34]{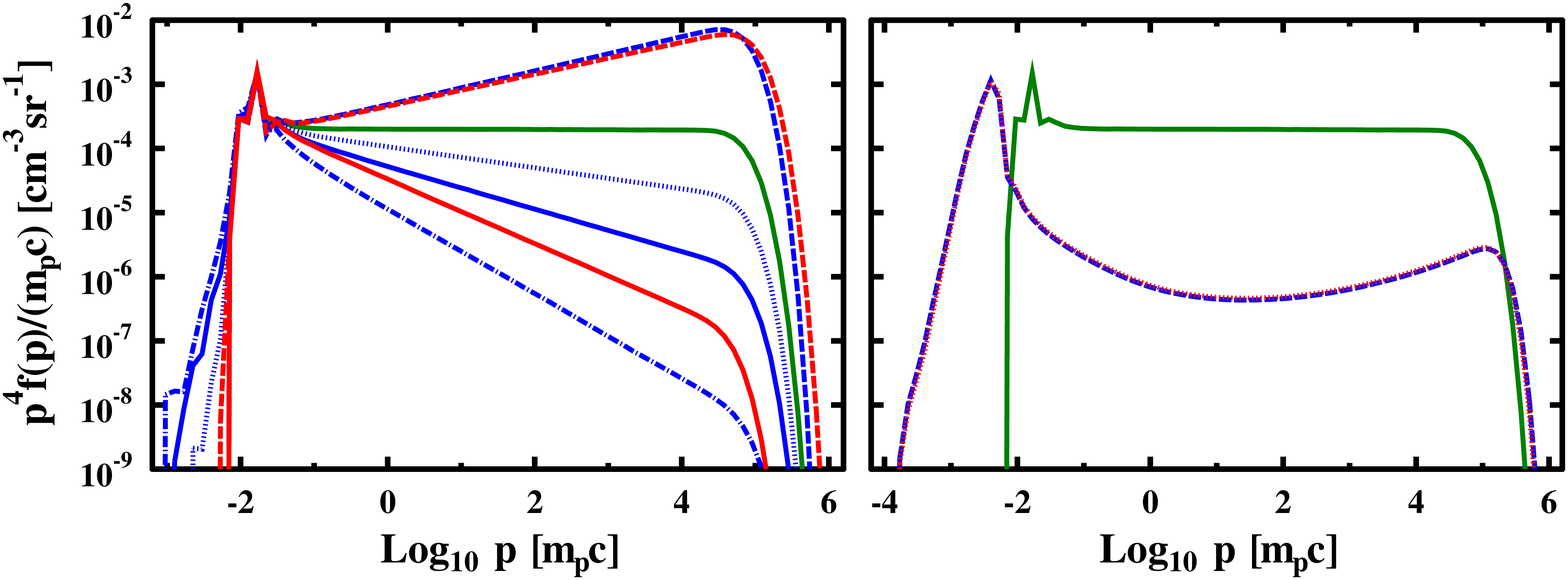}
\end{center}
\caption{Downstream shock frame proton distribution functions calculated with $\ush=5000$\,\kmps\ and $n_0=0.3$\,\pcc. In the left panel are test-particle results with $\Rtot=4$. The red curves ($E_0=0$): solid $\vscat=-\ush/4$ ($\alpha=4.5$), dashed $\vscat=+\ush/4$ ($\alpha=3.75$). The blue curves ($\vscat=0$): dot-dashed $E_0=2 \Eest$, solid $E_0=\Eest$, dotted $E_0=0.5\Eest$, dashed $E_0=-\Eest$.  $\Eest \approx 1.25\xx{-8}$ (cgs  units) for $\vscat=-\ush/4$ (Eq.~\ref{eq:Eest}).  In both panels, the green curve is a test-particle result with $\vscat=E_0=0$. In the right panel self-consistent nonlinear results with
$\vscat(x) \neq 0$ (red curve) and $E_0(x) \neq 0$
(blue curve) are presented.}
\label{pdf_sf}
\end{figure}

The particle spectra obtained with our test-particle calculations are shown in the left panel of Fig.~\ref{pdf_sf}.
In the model with $\vscat \neq 0$ and $E_{0}=0$
the distribution function at superthermal energies is a power-law
$f_0(p) \sim p^{-\alpha}$,
where
$\alpha = 3 \Reff/(\Reff -1)$ and $\Reff$ is the shock compression ratio, i.e., $\Reff = \usc^u/\usc^d$.
The subscript on $\Reff$ indicates ``effective".

\begin{table}[h]
\caption{\label{Tabel_1}Results of the self-consistence calculations.}
\begin{center}
\lineup
\begin{tabular}{*{5}{l}}
\br
$\ush$ [\kmps]&$n_{0}$ [\pcc]&$\0 \Rtot$&\m$\Beffd$[mG]&
\m$\varphi$ [TeV]\cr
\mr
\0\0\0 5000  &\0\0 0.3  & 9.4 (9.2)  &\0 0.63 (0.61)  &\0\0\0 2.4  \cr
\0\0\0 5000  &\0\0 1.0  & 8.9 (8.7)  &\0\0 1.2 (1.1)  &\0\0\0 3.9  \cr
\0\0 10000   &\0\0 0.3  & 7.8 (7.7)  &\0\0 1.4 (1.3)  &\0\0 10.7   \cr
\0\0 10000   &\0\0 1.0  & 7.5 (7.4)  &\0\0 2.6 (2.5)  &\0\0 15.8   \cr
\0\0 20000   &\0\0 0.3  & 6.7 (6.7)  &\0\0 3.1 (2.8)  &\0\0 37.6   \cr
\0\0 20000   &\0\0 1.0  & 6.6 (6.6)  &\0\0 5.5 (5.1)  &\0\0 51.6   \cr
\br
\end{tabular}
\end{center}
\end{table}

\begin{figure}[h]
\begin{center}
\includegraphics[scale=0.54]{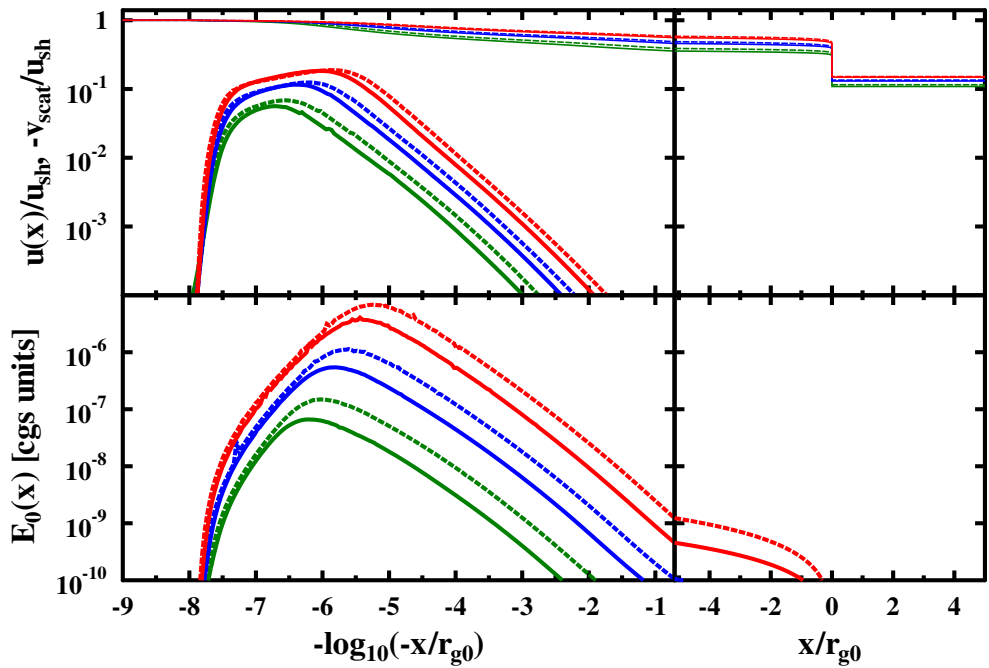}
\end{center}
\caption{Spatial profiles of the background plasma speed $u(x)$
(thin lines, top panel), 
the scattering center speed relative to the
background plasma $\vscat(x)$
(thick lines, top panel), and the average electric field $E_0(x)$ (bottom panel) for self-consistence calculations.
The curve color corresponds to shock speeds $\ush$:
$5000$\,\kmps\ - green line; $10^4$\,\kmps\ - blue line; and
$2\xx{4}$\,\kmps\ - red line. The solid lines for $n_{0}=0.3$\,\pcc, while the dashed lines for $n_{0}=1$\,\pcc.}
\label{plot_u_v_scat_E0}
\end{figure}

For our self-consistent, nonlinear models, we use two values for the background ambient number density, $n_0$, and three values  for the shock speed,
$\ush$ (see Table 1). All other parameters are the same except for the two closure models indicated by eqs.~(\ref{eq_v_scat}) and (\ref{eq_E0}).
The parameters are typical for forward shocks in SNRs interacting with the interstellar medium.
As indicated by the red and blue curves
in the right panel of Fig.~\ref{pdf_sf}, these two closure models give nearly identical results for our nonlinear shocks. 
The nonlinear profiles for  $E_0(x)$ and $\vscat(x)$
are presented in Fig.~\ref{plot_u_v_scat_E0}.
Other quantities are given in Table~\ref{Tabel_1} [values in parentheses indicate $\vscat(x)\neq 0$],
where $\Rtot$ is the full compression,
$\Beffd$ is the effective downstream field,
and $\varphi$ is the energy a proton acquires by traversing the upstream region through the average electric field, i.e., $\varphi= e \int_{\xFEB}^{0} E_0(x)dx$.

\section{Summary}
The self-generation of magnetic turbulence is an essential part of nonlinear Fermi acceleration. However, the production and amplification of turbulence is a complicated process, with an extremely  wide dynamic range if SNRs are to be modeled. While \PIC\ simulations give important information on small scales, none are yet large enough to give a consistent picture of Fermi acceleration and MFA for shocks typical of those seen in young SNRs.
With important approximations, the Monte Carlo techniques we
discuss can model critical aspects of nonlinear Fermi acceleration in shocks with large dynamic ranges. We have generalized previous results to include descriptions of the speed of scattering centers through the self-generated turbulence and the dynamic effects of non-zero local electric fields generated by Bell's instability.

While further study is required to determine  which, if either,
of these effects dominate the Fermi acceleration of ions, we have shown that both processes, expressed in Eqs.~(\ref{eq_v_scat}) and (\ref{eq_E0}), influence the accelerated proton spectrum in very similar ways
(right panel in Fig.~\ref{pdf_sf}).
It is also important to recognize  that radiation from SNRs can't be modeled properly unless electrons are included consistently
(see,  for example,  \cite{LSENP2013}).
As noted by \cite{Zirakashvili2015}, the fact that electrons have the opposite sign from protons may be important for the processes discussed here.
However, since electrons carry much  less momentum than protons, they will have little influence on the shock dynamics and we don't include them in our simulations.

In general, an average electric field $E_0(x) >0$ removes energy from protons and a scattering center speed $\vscat(x)<0$, i.e., one where the scattering centers move away from the viscous subshock in the upstream plasma frame,  results in a weaker effective compression ratio.
While both of these effects tend to soften the proton spectrum, the nonlinear shock modification, along with the self-consistent injection included in the Monte Carlo model, still produce a concave spectral shape with a slope harder than $p^{-4}$ before the cutoff at the highest energies (red and blue curves in the right panel of Fig.~\ref{plot_u_v_scat_E0}).
We note that $E_0(x)>0$ will tend to harden an electron spectrum while $\vscat(x)<0$ will  tend to soften spectra regardless of sign.


\ack
The authors acknowledge support from RAS Presidium program No. 12.
The results of this work were obtained using computational resources of the Peter the Great Saint-Petersburg Polytechnic University Supercomputing Center (http://www.spbstu.ru).




\section*{References}

\bibliographystyle{iopart-num}
\bibliography{bibliogr}

\providecommand{\newblock}{}
\begin{thebibliography}{10}
\expandafter\ifx\csname url\endcsname\relax
  \def\url#1{{\tt #1}}\fi
\expandafter\ifx\csname urlprefix\endcsname\relax\def\urlprefix{URL }\fi
\providecommand{\eprint}[2][]{\url{#2}}

\bibitem{Ellison12}
{Ellison} D~C, {Slane} P, {Patnaude} D~J and {Bykov} A~M 2012 {\em \apj\/} {\bf
  744} 39 (\textit{Preprint} \eprint{1109.0874})

\bibitem{VBK2005a}
{V{\"o}lk} H~J, {Berezhko} E~G and {Ksenofontov} L~T 2005a {\em \aap\/} {\bf
  433} 229--240

\bibitem{Jones91}
{Jones} F~C and {Ellison} D~C 1991 {\em \ssr\/} {\bf 58} 259--346

\bibitem{Bell15}
{Bell} A~R 2014 {\em Brazilian Journal of Physics\/} {\bf 44} 415--425
  (\textit{Preprint} \eprint{1311.5779})

\bibitem{Caprioli2010}
{Caprioli} D, {Amato} E and {Blasi} P 2010 {\em Astroparticle Physics\/} {\bf
  33} 307--311 (\textit{Preprint} \eprint{0912.2714})

\bibitem{SchureEtal2012}
{Schure} K~M, {Bell} A~R, {O'C Drury} L and {Bykov} A~M 2012 {\em \ssr\/} {\bf
  173} 491--519 (\textit{Preprint} \eprint{1203.1637})

\bibitem{BEOV14}
{Bykov} A~M, {Ellison} D~C, {Osipov} S~M and {Vladimirov} A~E 2014 {\em ApJ\/}
  {\bf 789} 137 (\textit{Preprint} \eprint{1406.0084})

\bibitem{Skilling1975}
{Skilling} J 1975 {\em \mnras\/} {\bf 172} 557--566

\bibitem{McKVlk82}
{McKenzie} J~F and {Voelk} H~J 1982 {\em \aap\/} {\bf 116} 191--200

\bibitem{Bell04}
{Bell} A~R 2004 {\em \mnras\/} {\bf 353} 550--558

\bibitem{Zirakashvili2008}
{Zirakashvili} V~N, {Ptuskin} V~S and {V{\"o}lk} H~J 2008 {\em \apj\/} {\bf
  678} 255--261 (\textit{Preprint} \eprint{0801.4486})

\bibitem{Zirakashvili2015}
{Zirakashvili} V~N and {Ptuskin} V~S 2015 {\em Bulletin of the Russian Academy
  of Sciences, Physics\/} {\bf 79} 316--318

\bibitem{Fedorov1992}
{Fedorov} Y~I, {Kats} M~F, {Kichatinov} L~L and {Stehlik} M 1992 {\em \aap\/}
  {\bf 260} 499--509

\bibitem{Vladimirov09dis}
{Vladimirov} A 2009 {\em {Modeling magnetic field amplification in nonlinear
  diffusive shock acceleration}\/} Ph.D. thesis North Carolina State University

\bibitem{EBJ96}
{Ellison} D~C, {Baring} M~G and {Jones} F~C 1996 {\em \apj\/} {\bf 473} 1029--+

\bibitem{Bykov11}
{Bykov} A~M, {Osipov} S~M and {Ellison} D~C 2011 {\em \mnras\/} {\bf 410}
  39--52 (\textit{Preprint} \eprint{1010.0408})

\bibitem{bk88}
{Berezhko} E~G and {Krymski{\u i}} G~F 1988 {\em Soviet Physics Uspekhi\/} {\bf
  31} 27--51

\bibitem{LSENP2013}
{Lee} S~H, {Slane} P~O, {Ellison} D~C, {Nagataki} S and {Patnaude} D~J 2013
  {\em \apj\/} {\bf 767} 20 (\textit{Preprint} \eprint{1302.4645})

\end{thebibliography}
\end{document}